\documentclass[a4paper]{jpconf} 
\usepackage{graphicx}

\usepackage{amssymb}

\begin{document}
\title{Statistical Analysis of weather variables of Antofagasta}

\author{H. Farfan$^\dag {}^1$, S. Curilef${}^1$ and A. Castillo${}^2$}

\address{${}^1$Departamento de F\'isica, Universidad Cat\'olica del Norte, Antofagasta, Chile.}
\address{${}^2$Laboratorio de Sedimentolog\'ia y Paleoambientes, Universidad de
Antofagasta, Chile.}


\ead{$^\dag$hishan.farfan@ucn.cl}

\begin{abstract}
The statistical behavior of weather variables of Antofagasta is described, especially the daily data of air as temperature, pressure and relative humidity measured at 08:00, 14:00 and 20:00. In this article, we use a time series deseasonalization technique, Q-Q plot, skewness, kurtosis and the Pearson correlation coefficient. We found that the distributions of the records are symmetrical and have positive kurtosis, so they have heavy tails. In addition, the variables are highly autocorrelated, extending up to one year in the case of pressure and temperature.
\end{abstract}


\section{Introduction}
It is well-known that atmospheric systems essentially have a non-linear behavior due to a large number of factors that affect its large-scale evolution as the phenomenon of global atmospheric circulation \cite{makarieva2017} and small scale as the geographic conditions of the place. Antofagasta is a city located in the north of Chile and its stable climate is due to the almost permanent presence of the Southeast Pacific Subtropical Anticyclone (SPSA), the proximity of the Andes and the Humboldt current. The phenomenon of El Ni\~no-Southern Oscillation (ENSO) is also an important factor in the dynamics of the atmosphere \cite{garreaud2011}.
The main objective of this work is to study the statistical behavior of the atmospheric variables of Antofagasta inspecting the distribution of the data with a Q-Q plot, the skewness and kurtosis, and finally and by the use of the Pearson correlation coefficient to measure autocorrelation.
The data used in this study were obtained from the meteorological station in Universidad Cat\'olica del Norte (23.4$^\circ$S, 70.2$^\circ$W), located at 31 meters above sea level, whose daily records of temperature, atmospheric pressure and relative humidity measured since 1969 till 2016 were taken into account.
\newpage
\section{Metodology}\label{Section:Metodologia}

\subsection{Deseasonalization}\label{Subsection:Desestacionalizacion}
To remove the seasonal cycle of a time series, the \textit{daily means} is calculated, which is the average of the values sharing the same day in the year. If each corresponding value of the original series is subtracted from its corresponding daily average, the resulting values are called \textit{residues}. The calculation of deseasonalization is obtained considering $\tilde{X}$ as the original time series of size \textit{N} and subtracting the daily averages to each corresponding day \textit{d}, that is, $<\tilde{X}>_d$, obtaining the residuals $X$.
\begin{equation}\label{Ecuacion:Desestacionalizacion}
X=\tilde{X}_d-\left<\tilde{X}\right>_d
\end{equation}

\subsection{Q-Q plot}\label{Subsection:GraficoQ-Q}
\begin{figure}[h]
\begin{minipage}{15pc} 
\includegraphics[width=30pc]{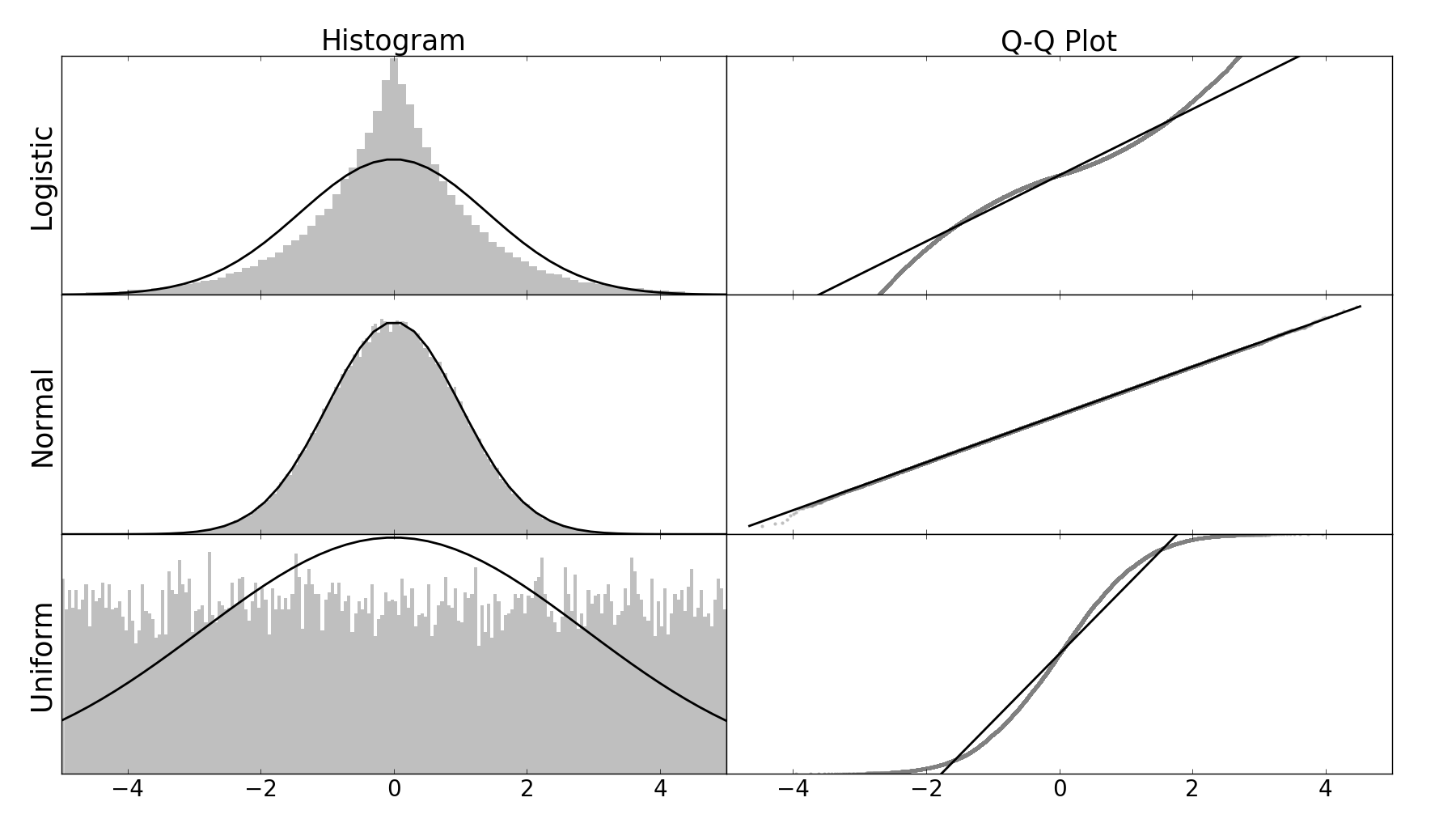}
\end{minipage}\hspace{15pc} 
\begin{minipage}{8pc} 
\caption{Histograms with normal fit and normal Q-Q plot of logistic distribution (top), normal distribution (center) and uniform distribution (bottom).}
\end{minipage} 
\label{Figura:QQPlot}
\end{figure}

It is a graphical method to compare the probability distribution of a sample to a theoretical distribution of interest (usually normal) by plotting the quantiles, which are intervals divide a distribution into equal portions, one distribution against the other. If we plot the sample data against the generated values, a straight line with slope 1 is formed, then both values come from the same probability distribution \cite{wilk1968}. As an example, figure \ref{Figura:QQPlot} shows the case of three different types of distributions and their respective Q-Q plots. In the case of the logistic distribution, it can be seen how the values close to the average are underestimated by the normal distribution, and the values away from the center they are overestimated, whereas for the uniform distribution it is the other way around, with the central values being overestimated and the extremes underestimated.

\subsection{Skewness and Kurtosis}
The mean ($\mu$) and the variance ($\sigma^2 $) are quantities usually used to describe the form of a probability distribution, which in turn are derived from a mathematical concept called \textit{moment}, the first being a central measure of the data and the second a measure of the average dispersion around the mean. However, this concept can be generalized until obtaining moments of greater order, such as \textit{skewness} and \textit{kurtosis} \cite{kokoska2000}, where the first is a measure of symmetry of the distribution around the mean and the second, a measure of the way in which the tails of the probability distribution fall (see Table \ref{Cuadro:Momentos}).
\begin{center}
\begin{table}[!htpb]
\caption{Moments of a probability distribution}
\centering
\begin{tabular}{@{}l*{7}{l}}
\br
n & Name & Symbol & Expression & Standard normal distribution\\
\mr
1 & Mean & $\mu$ & $\mathbb{E}(x)$ & \hspace{2cm} 0\\
2 & Variance & $\sigma^2$ & $\mathbb{E}\left[\left(X-\mu\right)^2\right]$ & \hspace{2cm} 1\\
3 & Skewness & $\gamma_1$ & $\mathbb{E}\left[\left(\frac{X-\mu}{\sigma}\right)^3\right]$ & \hspace{2cm} 0\\
4 & Kurtosis & $\gamma_2$ & $\mathbb{E}\left[\left(\frac{X-\mu}{\sigma}\right)^4\right]$ & \hspace{2cm} 3\\
\br
\end{tabular}\label{Cuadro:Momentos}
\end{table}
\end{center}
In the case of skewness, if after its calculation a positive value is obtained, it is said to have positive skewness and the upper tail falls more slowly than the lower one, otherwise, the skewness is negative. As for kurtosis, if a positive value is obtained, the distribution is called \textit{Leptokurtic} where the tails fall quickly as in the case of the logistic distribution, otherwise if the kurtosis is negative, it is called \textit{Platykurtic}, where its behavior is similar to the uniform distribution (see figure \ref{Figura:QQPlot}).
\subsection{Correlation coefficient}
The Pearson correlation coefficient is obtained by dividing the covariance of two variables between the product of its standard deviations and calculated as  \cite{cohen2014}.
\begin{equation}\label{Ecuacion:Correlacion}
r=\frac{\sum_{i=1}^N (x_i-\mu_x)\cdot(y_i-\mu_y)}{\sqrt{\sum_{i=1}^N (x_i-\mu_x)^2}\cdot \sqrt{\sum_{i=1}^N (y_i-\mu_y)^2}},
\end{equation}
and for two series with a certain offset $ k $ we get
\begin{equation}\label{Ecuacion:Autocorrelacion}
r_k=\frac{\sum_{i=1}^{N-k} (x_i-\mu_x)\cdot(y_{i+k}-\mu_y)}{\sqrt{\sum_{i=1}^{N-k} (x_i-\mu_x)^2}\cdot \sqrt{\sum_{i=k+1}^N (y_i-\mu_y)^2}}.
\end{equation}
This coefficient is 1 for two series with perfect positive linear correlation, 0 for statistically independent series and -1 for series with perfect negative linear correlation.

\section{Results}
We start by obtaining the \textit{residuals} from the original data shown in figure 2 with the deseasonalization method described in equation (\ref{Ecuacion:Desestacionalizacion}). Observing the graph of the residual data measured at 08:00, which are shown in figure \ref{Figura:SerieTiempo_R}, the fluctuations of the pressure, temperature and relative humidity are evident. At this point, we can see two areas where it is distinguished that the pressure slightly decreases and the temperature considerably increases, these are the ENSO 82-83 and ENSO 97-98.\newpage
\begin{figure}[h]
\begin{minipage}{15pc} 
\includegraphics[width=30pc]{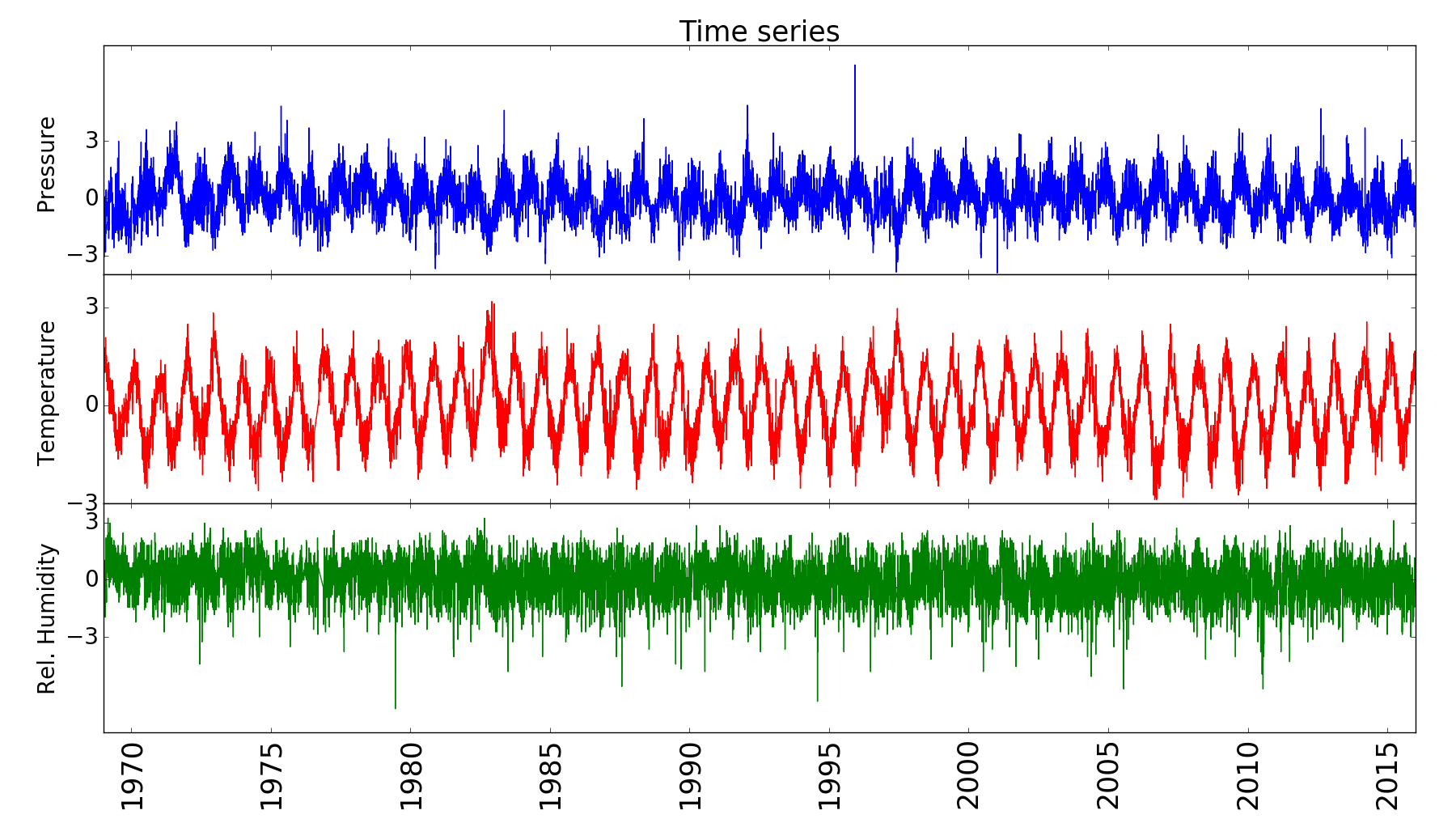}
\end{minipage}\hspace{15pc} 
\begin{minipage}{8pc} 
\caption{Pressure (top), temperature (centro) and relative humidity (bottom) measured at 08:00.}
\end{minipage} 
\label{Figura:SerieTiempo_AA}
\end{figure}
\begin{figure}[h]
\begin{minipage}{15pc} 
\includegraphics[width=30pc]{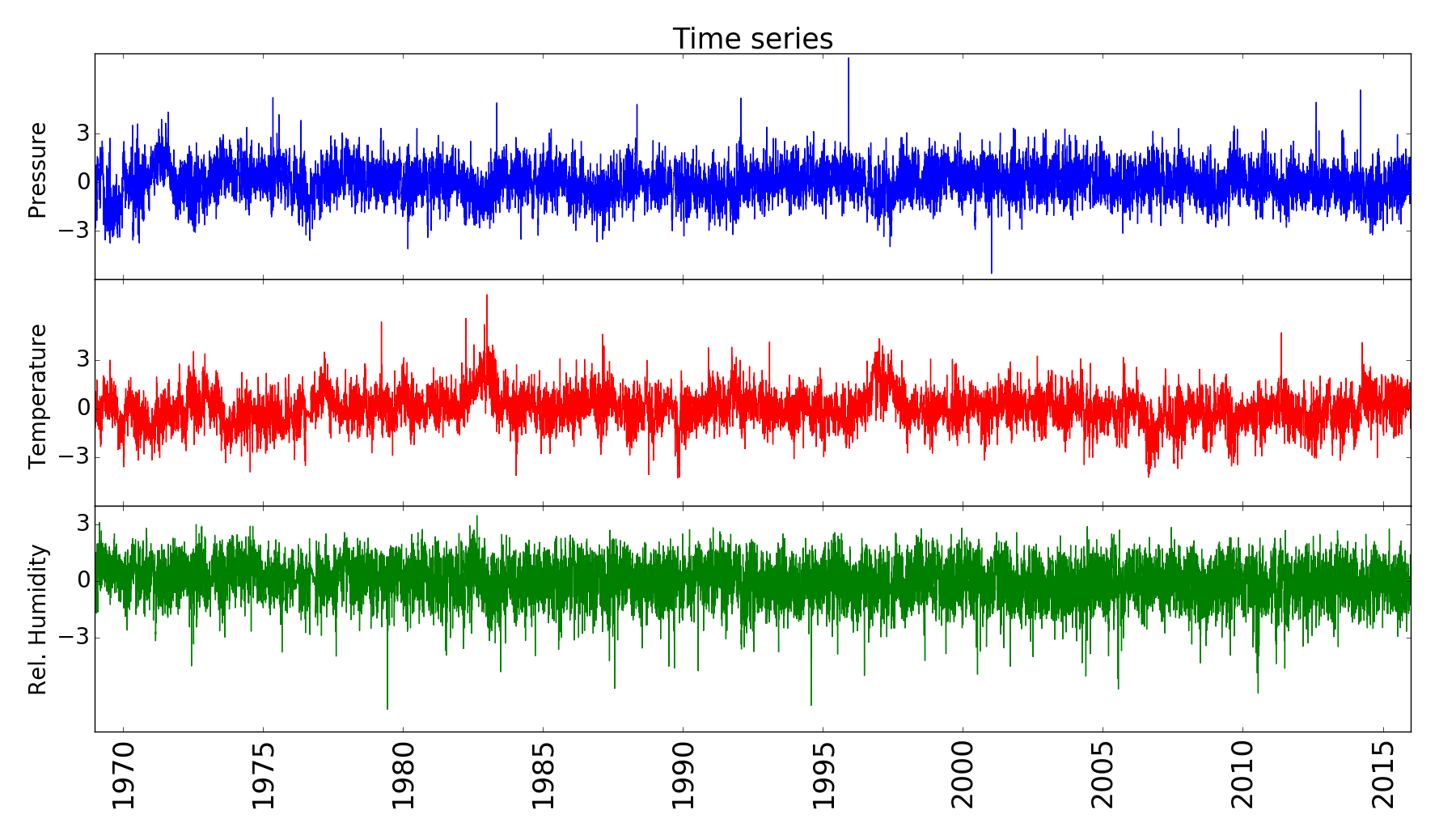}
\end{minipage}\hspace{15pc} 
\begin{minipage}{8pc} 
\caption{Residuals of pressure (top), temperature (centro) and relative humidity (bottom) measured at 08:00.}
\end{minipage} 
\label{Figura:SerieTiempo_R}
\end{figure}

\subsection{Probability distribution and Q-Q plot}
\begin{figure}[h]
\begin{minipage}{15pc} 
\includegraphics[width=30pc]{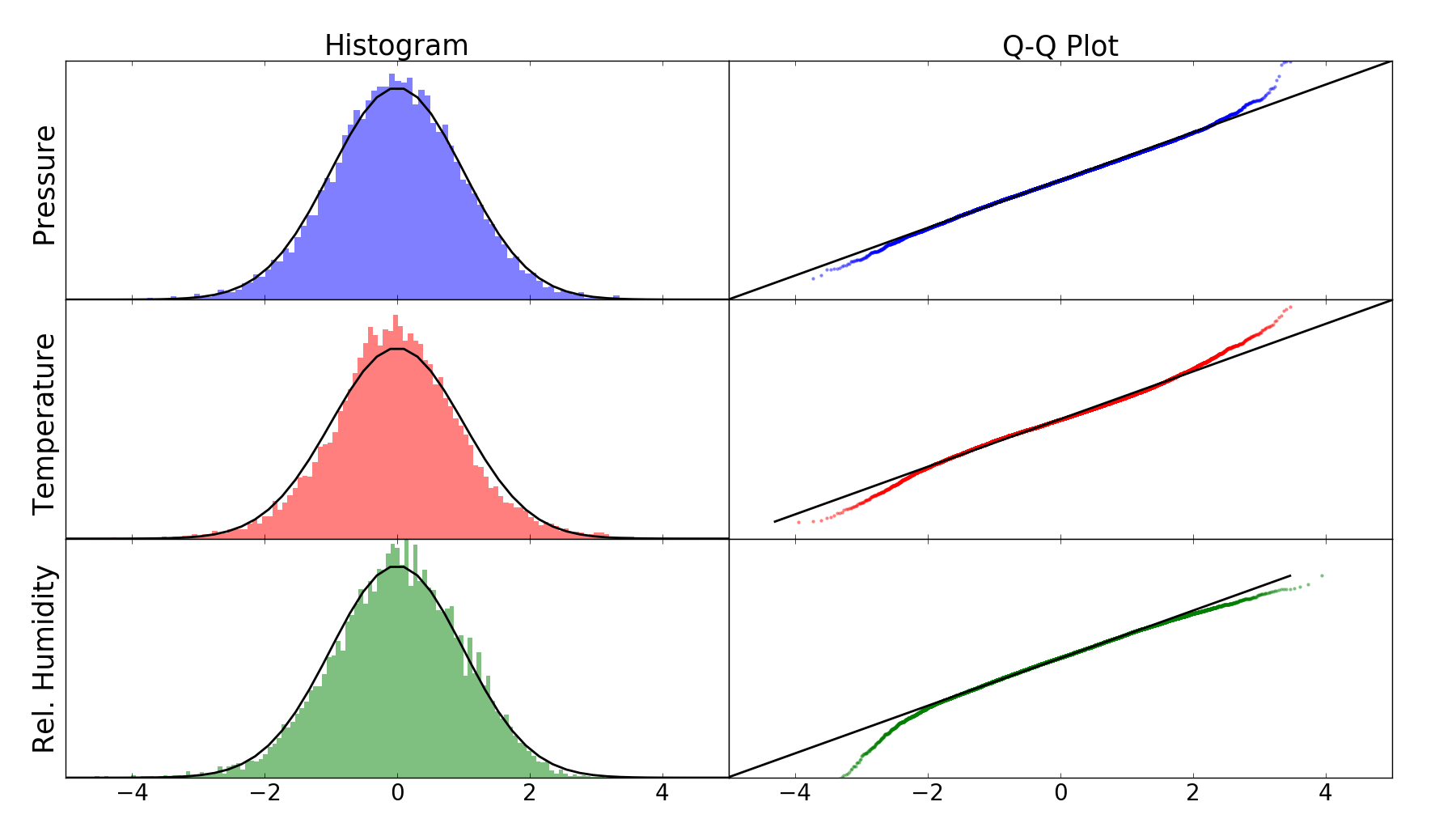}
\end{minipage}\hspace{15pc} 
\begin{minipage}{8pc} 
\caption{Probability distribution and Q-Q plot of pressure (top), temperature (center) and relative humidity (bottom), measured at 08:00.}
\end{minipage} 
\label{Figura:QQPlot_R}
\end{figure}

In the left part of the figure \ref{Figura:QQPlot_R} the probability distributions of the residuals of the pressure, temperature and relative humidity measured at 08:00 are shown, whose shapes seem to be normal distributions. On the right side, the Q-Q plot is shown as defined in the section \ref{Subsection:GraficoQ-Q}, where the distribution of the three variables fits the theoretical normal distribution quite well, except for the values farthest from the mean. The fall of the tails of the distributions of the pressure  and the temperature drops quickly after moving away from the average and then falling slowly as they move away from the center, as in the case of the logistic distribution (see figure \ref{Figura:QQPlot}).
This is corroborated by the skewness ($\gamma_1 $) and kurtosis ($\gamma_2 $) values (see table \ref{Cuadro:Momentos}) of each variable shown in the table \ref{Cuadro:AsimetriaYCurtosis}. The value close to zero of the skewness of the variables is reflected in the symmetry of the distributions around the mean, on the other hand the excessively positive kurtosis (greater than 3) supports the anomalous behavior in the fall of the tails in a non-exponential way, which are commonly known as ``heavy tails'' \cite{asmussen2003}, and suggests memory in the records.
\begin{center}
\begin{table}[!htpb]
\caption{Skewness ($\gamma_1$) and Kurtosis ($\gamma_2$) of pressure (PA), temperature (TT) and relative humidity (HR) measured at 08:00, 14:00 and 20:00.}
\centering
\begin{tabular}{@{}l*{9}{l}}
\br
& PA08 & PA14 & PA20 & TT08 & TT14 & TT20 & HR08 & HR14 & HR20\\
\mr
$\gamma_1$ & 0.05 & -0.02 & 0.06 & 0.11 & -0.23 & 0.44 & -0.39 & 0.02 & -0.27 \\
$\gamma_2$ & 3.76 & 3.47 & 3.48 & 4.14 & 8.37 & 4.36 & 3.96 & 3.73 & 3.69 \\
\br
\end{tabular}\label{Cuadro:AsimetriaYCurtosis}
\end{table}
\end{center}

\subsection{Autocorrelation}
\begin{figure}[h]
\begin{minipage}{15pc} 
\includegraphics[width=30pc]{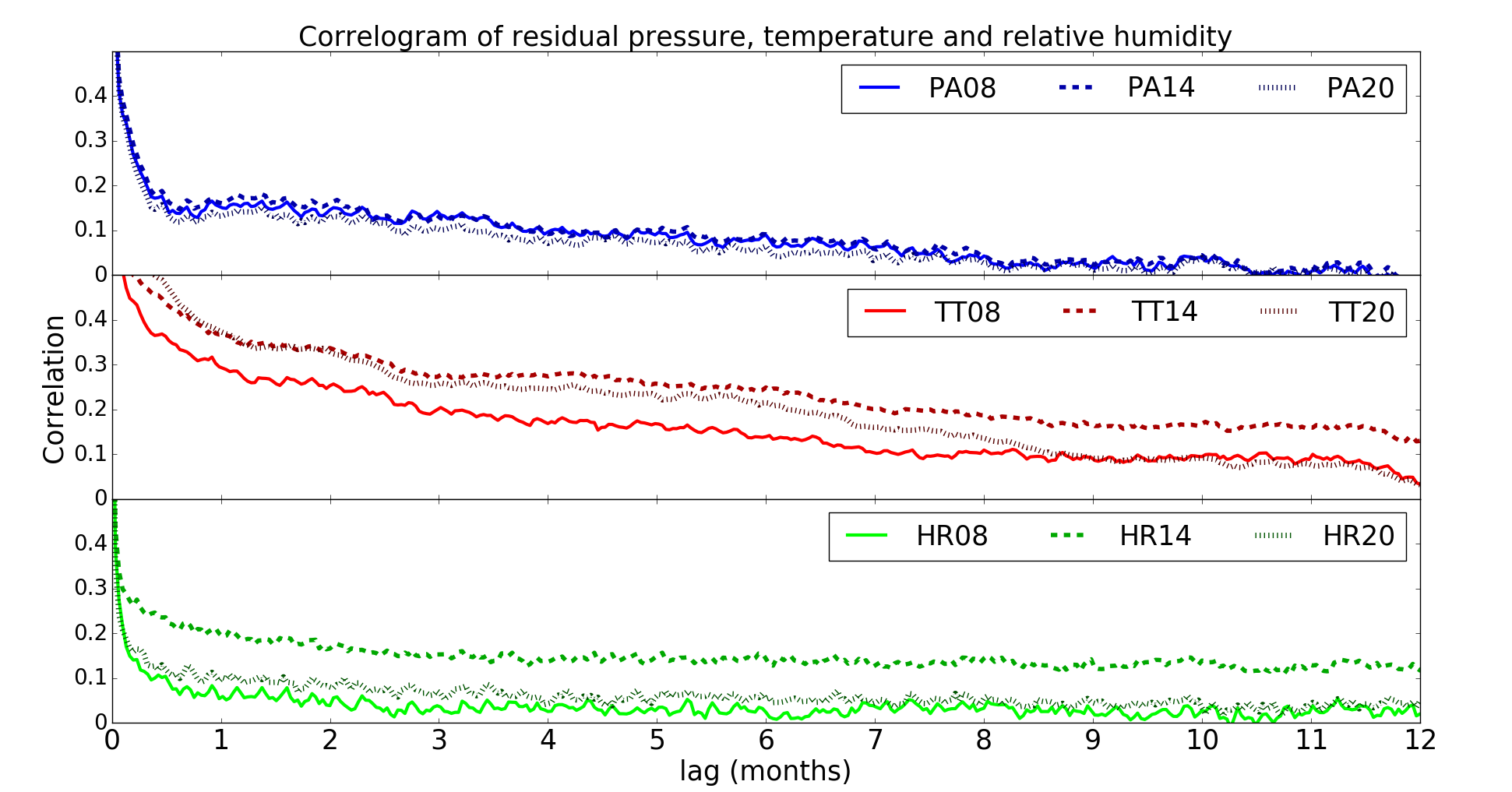}
\end{minipage}\hspace{15pc} 
\begin{minipage}{8pc} 
\caption{Correlogram of pressure (top), temperature (center) and relative humidity (bottom).}
\end{minipage} 
\label{Figura:Correlograma_PA_TT_HR}
\end{figure}

Using equation (\ref{Ecuacion:Autocorrelacion}), the correlogram for the variables and the three hours of measurement is shown in figure \ref{Figura:Correlograma_PA_TT_HR}. Autocorrelation behavior of the variables are different among them; nevertherless, for each variable, the behavior itself is similar at all hours. The autocorrelation of the pressure drops abruptly until approximately one month, then slowly decreases to almost zero after approximately 10 months. For temperature, the fall is even slower, whose values almost vanishes after about a year, which highlights the measurement at 14:00, where after this period the autocorrelation barely falls to be weak but not zero. In the case of relative humidity, a fast fall is observed after a few weeks, then stabilize until extremely slow for 08:00 and 20:00, and in the case of 14:00, the fall after approximately one month is imperceptible, where after a year, the autocorrelation is weak but still persistent.

\section{Discussion and Summary}
The pressure and the temperature and, to a lesser extent, the relative humidity, have a strong annual pattern (see figure \ref{Figura:SerieTiempo_AA}), which we must remove to study the variables over time regardless of the time of year. All distributions are symmetric in relation to the mean, but they have positive kurtosis (see table \ref{Cuadro:AsimetriaYCurtosis} and figure \ref{Figura:QQPlot_R}), which indicates that the variables are autocorrelated, as corroborated with the analysis of correlation.
The main variables have shown correlations that are persistent for several months. For the pressure and the temperature last until the end of the year. For the relative humidity, the correlation remains upto one month. In addition, the pressure and the temperature endure autocorrelated for a long time, while the solar radiation is high, this is, at 14:00 (see figure \ref{Figura:Correlograma_PA_TT_HR}).

\ack
We would like to thank partial financial support from FONDECYT 1181558. H.F. is really grateful to Francisco Calder\'on for the assistance in the computer programming.

\section*{References}
\bibliographystyle{ieeetr}
\bibliography{bibliography}{} 

\end{document}